# Observation-based Development of Software Process Baselines: An Experience Report

Fabio Bella, Jürgen Münch, and Alexis Ocampo

***Abstract*— The creation and deployment of software development processes for new domains (such as wireless Internet services) is a challenging task due to the lack of knowledge about adequate engineering techniques and their effects. In addition, time-to-market pressure prevents applying long-lasting maturation of processes. Nevertheless, developing software of a predetermined quality in a predictable fashion can only be achieved with systematic development processes and the use of engineering principles. A descriptive approach promises to quickly create initial valuable process models and quantitative baselines that can be seen as starting points for continuous improvement activities.**

**This paper describes the creation of software development processes for the development of wireless Internet services based on the observation of pilot projects that were performed at distributed development sites. Different techniques and tools such as descriptive process modeling, process documentation generation, goal-oriented measurement, and capturing qualitative experience were combined to gain process baselines. Results show that the observation-based approach helped to quickly come up with stable context-oriented development processes and get a better understanding of their effects with respect to quantitative and qualitative means.**

## I. Introduction

NEW and therefore unknown application domains are often characterized by a lack of explicitly defined software development processes. Results from studies such as the multi-case study conducted at Nokia Mobile Phones clearly show the importance of a stable and well-implemented infrastructure for process deployment [15]. In the case of new domains, the design and introduction of processes can be very risky, since no previous experience exists on which processes or process fragments are suitable and executable in the environment of the developing organization.

Manuscript received April 1, 2004. This work was funded by the European Commission in the context of the WISE project (No. IST-2000-30028).

Fabio Bella, Dr. Jürgen Münch, and Alexis Ocampo are with the Fraunhofer Institute for Experimental Software Engineering, 67661 Kaiserslautern, Germany. Email: {bella, muench, ocampo}@iese.fraunhofer.de

The engineering of wireless Internet services shows exemplarily some of the difficulties inherent to the selection of appropriate processes in a new domain: If a specific process for wireless Internet services is not defined, the risk exists that the process followed in Internet services development will also be inherited for wireless services. As the wireless Internet gets popular, the Internet service providers will try to provide similar services over the wireless Internet as well, and they may easily try to follow the same development process they use for Internet services, although these processes do not address any of the problems specific to the wireless domain. An appropriate, domain-specific development process and process-related experience are needed very quickly.

This article describes a study for the observation-based creation of development processes for a new domain. The underlying approach can be applied to unknown new domains in general, but as the focus of this work is the wireless Internet domain, special emphasis is placed on the particularities of this domain.

In the following, Section 2 introduces the techniques applied within the study and explains how they relate to it. Section 3 discusses the context, in which the study was performed, and the overall approach applied to create process-related quality models. Concrete examples gathered from the observation of the development of representative pilot services present the approach from a practical point of view. Section 4 subsumes the article.

## II. Background

The techniques applied for project tracking and analysis within the scope of this article are a combination of descriptive process modeling [4], GQM-based measurement [5], and retrospective-based collection of lessons learned [7]. This section gives an overview of the methodologies applied and explains how they relate to the study.

The main idea of descriptive process modeling is to explicitly document the development processes as they are applied within a given organization: A so-called process engineer observes, describes, and analyzes the software development process and its related activities, and provides descriptions of the processes to the process performers. Since the processes are usually complex, support is needed for both process engineers and process performers. Descriptive process modeling is applied within the context of the study with the help of the Spearmint® environment. The architecture of Spearmint® and its features for a flexible definition of views, used for retrieving filtered and tailored presentations of process models, is presented in [4]. One distinct view, namely the Electronic Process Guide (EPG), is used for disseminating process information and guiding process performers. An EPG can be successfully applied in a distributed development environment, since it is Web-based (see Figure 1).

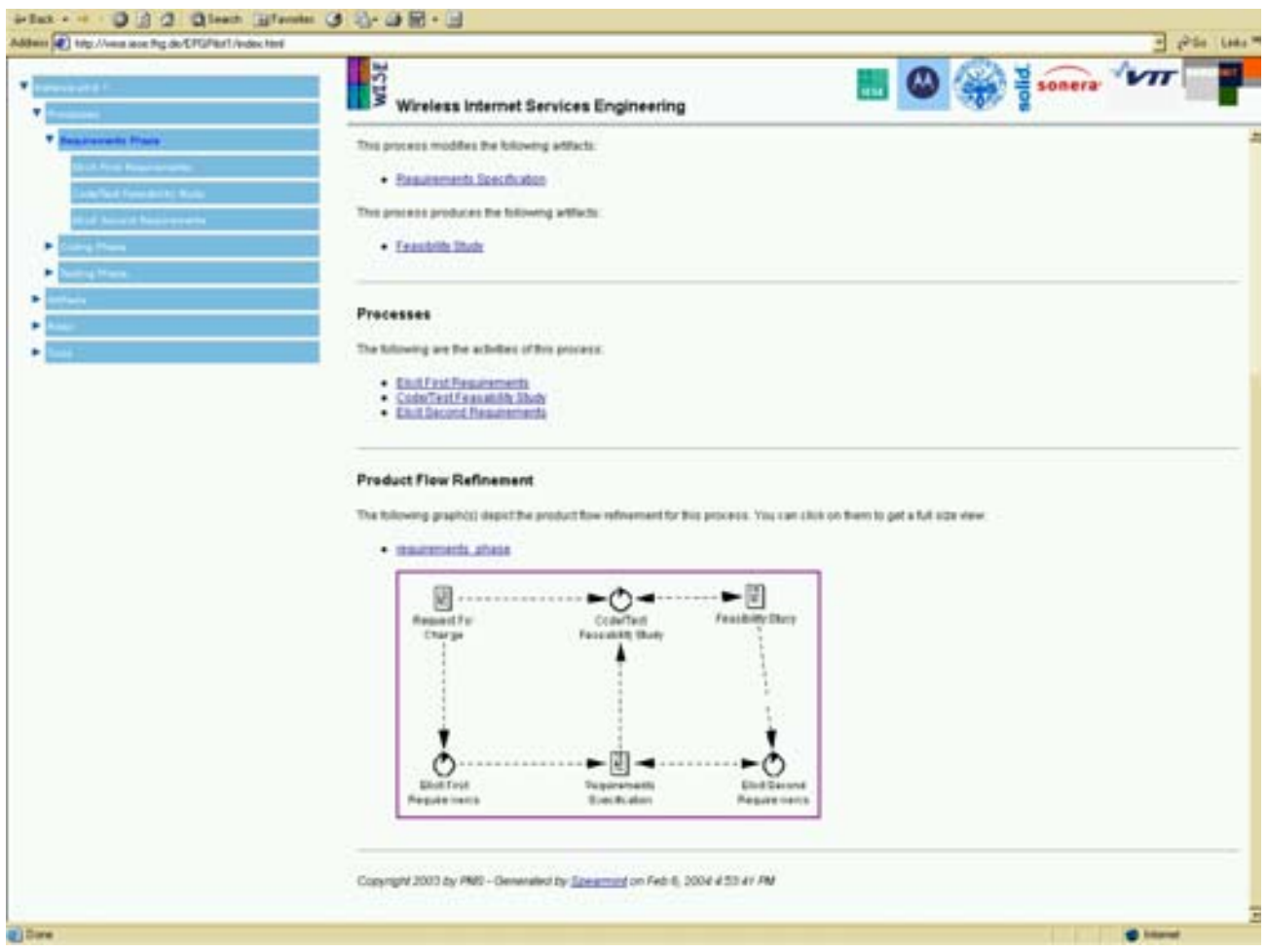

**Figure 1: Excerpt of an Electronic Process Guide**

The Goal/Question/Metric (GQM) approach is applied to collect process-related quantitative data. Briand et al. [5] describe this approach in terms of six major steps: During the first two steps, business and improvement goals are analyzed and metrics defined according to the process model that describes the whole development process as is; The results of this first phase are GQM plans that comprise all metrics defined. In the following step, the project plan and the process model are used to determine by whom, when, and how data are to be collected according to the metrics; The data collection procedures are the results of this instrumentation; Raw data are collected according to the data collection procedures; Quality models (baselines) are built and analyzed on the basis of the data collected and according to the GQM plans; In the fifth step, the interested parties interpret the baselines and draw possible consequences for the future; Finally, baselines, analysis, interpretations, and consequences are resumed and stored in the experience database for future reuse.

In addition to the measurement of quantitative data, the collection of qualitative data is driven by project retrospectives. Therefore, meetings and interviews with the participants of different work packages are conducted regularly. The key questions proposed by Kerth in [7] are addressed to focus the discussion on learning and improvement:

- What did we do well, which we might forget if we don't discuss it?
- What did we learn?
- What should we do differently next time?
- What still puzzles us?

## III. Data Collection and Analysis based on Explicit Process Models

In the following, the context of the study and the approach applied to create process-related baselines are described. Also, two concrete examples are given to show the approach from a practical point of view.

### A. *Project Context*

This work was conceived as an integral part of the evaluation of the Wireless Internet Service Engineering (WISE) project. The project aims at producing integrated methods and components (COTS and open source) to engineer services on the wireless Internet. The production of methods and components is driven by the development of pilot services.

The methods already produced include a reference architecture, a reference development process model, as well as guidelines for handling heterogeneous mobile devices.

The components include a service management component and an agent-based negotiation component.

Three pilot services, i.e., a financial information service, a multi-player game, and a data management service, are being developed by different organizations.

The project lasts for 30 months and an iterative, incremental development style is applied: three iterations are performed of roughly 9 months each.

In iteration 1, a first version of the planned pilot services was built using GPRS. At the same time, a first version of methods and tools was developed.

In iteration 2, a richer, second version of the pilots was developed on GPRS, using the first version of methods and tools. In parallel, an improved second version of methods and tools was developed.

In iteration 3, the final version of the pilots is being developed on UMTS, using methods and tools from the second iteration. Also, a final version of methods and tools is being developed.

The environment of the WISE project should be regarded as highly distributed, since several organizations from three different European countries are involved in it.

### B. *Stepwise Creation of Process Baselines*

This subsection presents the approach applied to track the development of the pilot services, acquire and analyze experience, and consequently improve the software processes followed within the WISE project.

As mentioned in the previous subsection, in parallel to the development of the pilot services, a measurement infrastructure was defined in order to evaluate the effects of the method and tools applied to develop these services. The infrastructure is based not only on measures but also on interviews and any other available evidence.

Figure 2 sketches the strategy applied iteratively during each of the three iterations to gather, package, and maintain experience from the development of the pilot services.

At the beginning of the iteration (set-up), software development processes are elicited as applied by the organizations; the descriptive process models and the measurement goals previously defined (mainly the characterization of effort, calendar time, and defects) are used to set up measurement programs, i.e., to define measurement plans.

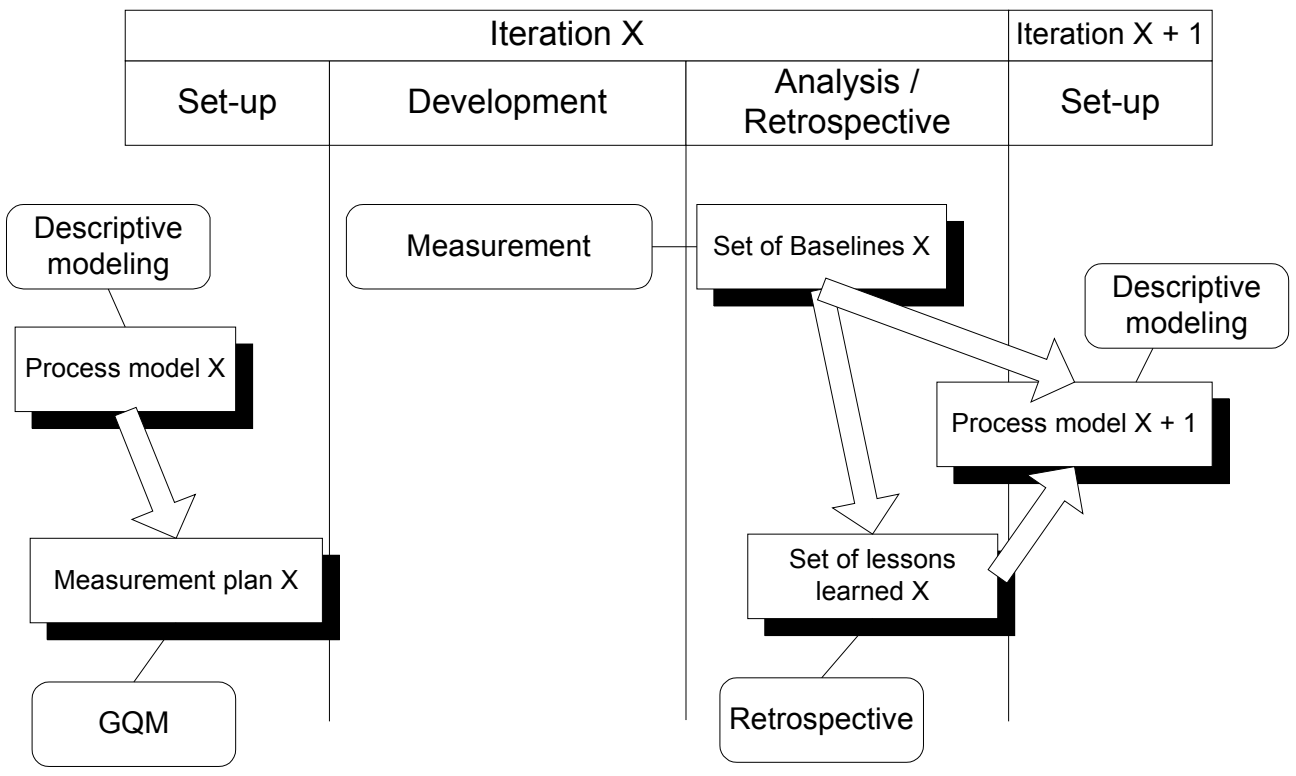

**Figure 2: Experience acquisition**

During the iteration (development), the pilot services are developed and the pilot performers collect data according to the measurement plans. The data is validated and stored.

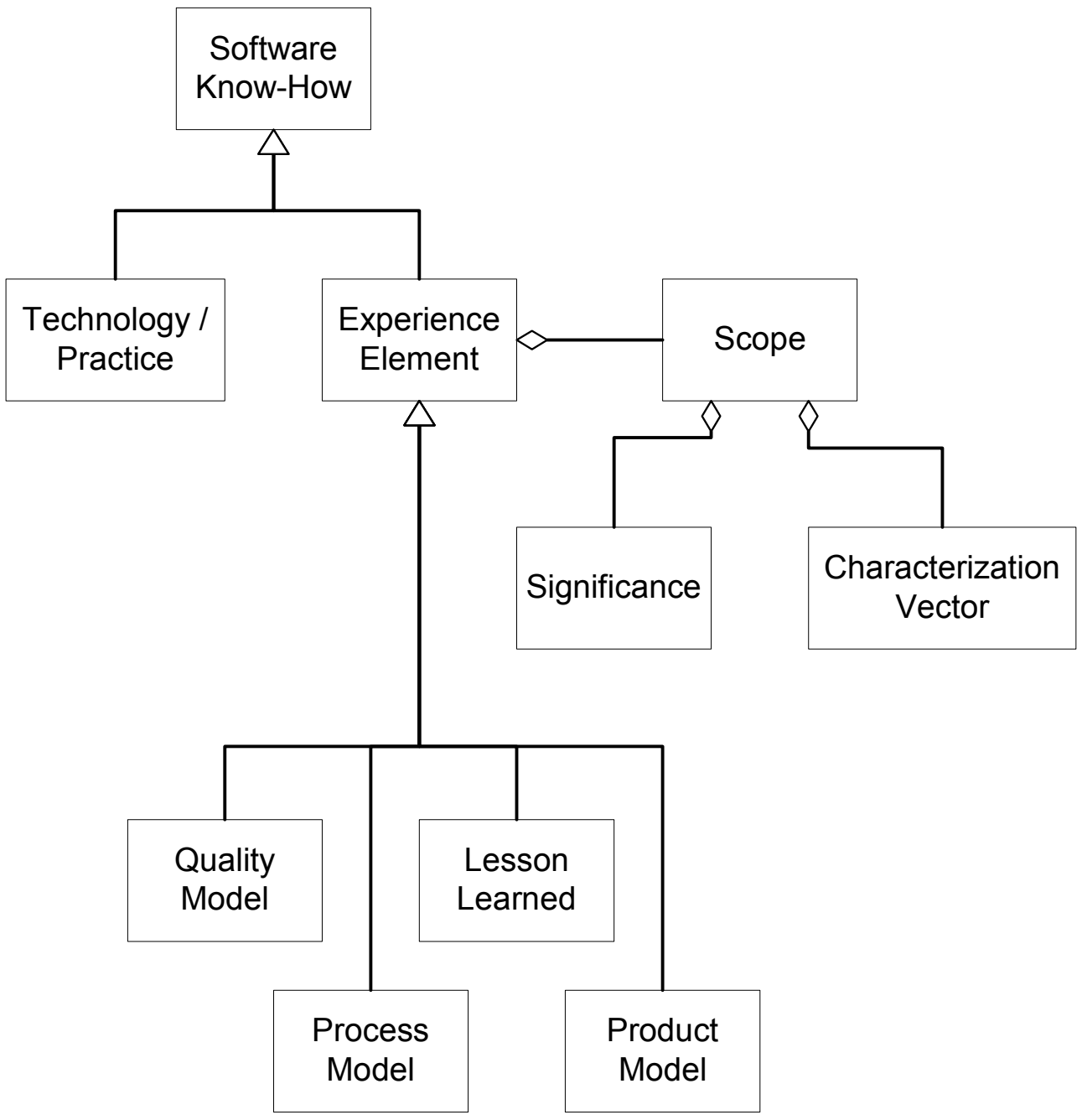

**Figure 3: Overview of experience elements**

At the end of the iteration (analysis / retrospective), baselines are built: the data collected are aggregated and analyzed together with the involved parties, then interpretations and consequences for the next iteration are worked out. In order to get more insights of a qualitative nature, lessons learned are collected regularly by interviewing project participants at project meetings or by phone. Many lessons were also gathered through the analysis and interpretation of baselines. Therefore, within the context of the WISE project, different experience models were applied (see Figure 3), which are an adaptation of basic principles of the Experience Factory [1] and QIP [2] approaches.

All kinds of software engineering experience are regarded as experience elements: process and product models, quantitative baselines, and qualitative experience such as lessons learned. For each experience element, the scope of its validity is described.

The scope consists of a characterization vector and the significance. The characterization vector characterizes the environment in which the experience element is valid (see Table 1). The significance describes how the experience element has been validated and to which extent (e.g., validation through formal experiments, single case study, or survey).

**Table 1:**
**Excerpt of a characterization vector**

| Customization factor | Characteristic | Pilot X |
|---|---|---|
| Domain characteristics | Application type | Computation-intensive system |
| | Business area | Mobile online entertainment services |
| Development characteristics | Project type | Client new development<br>Server new development |
| | Transport protocol | GSM/GPRS/UMTS |
| | Implementation language | Client: J2ME<br>Server: J2EE |
| | Role | Technology provider, service developer |

## C. *Results*

This subsection presents selected results achieved during the first two development iterations. Two concrete examples show how EPGs and quality models played a major role in learning lessons from the pilots and improving the process and the estimation capabilities of the organizations involved. The last subsection presents some lessons learned from developing the pilot services.

### 1) *Descriptive Process Creation*

Description of the pilot - Pilot service 1 provides a solution for real time stock tracking on mobile devices: the user can view real time quotes concerning a whole market or define his/her own watch lists. The partner responsible for this development is a provider of high end trading services on the Internet, aimed at banks and brokers. The pilot is the adaptation of an existing Web-based information service. Critical usability issues arise due to the huge amount of data needed by a financial operator to perform an analysis and the small size of the display of mobile devices. Furthermore, since the Internet traffic on mobile devices is paid for by the end user, based on data volume and not on connection time, and since frequent refresh of a large amount of financial data is required, the adoption of the push technology instead of the pull technology is an important issue, because it avoids unnecessary data refreshes for the user. Most of the usability issues were addressed during the first iteration. The second iteration was mainly concerned with implementing a solution based on the push technology.

The life cycle model applied for developing the pilot service during each iteration is an iterative process model consisting of three phases: a requirements phase, a development / coding phase, and a testing phase. The ad hoc process is characterized by extensive use of verbal communication within the development team, and little use of explicit documentation. Another important characteristic of the development process is the absence of an explicit design phase. This can be seen as a consequence of the fact that mainly the same client server architecture used to provide the service on the traditional Internet was applied: during the first iteration, the client side was a prototype developed using the Wireless Markup Language (WML); during the second iteration, the client was developed using the Java 2 platform, Micro Edition (J2ME). In both cases, the prototype and its high level design were documented after development.

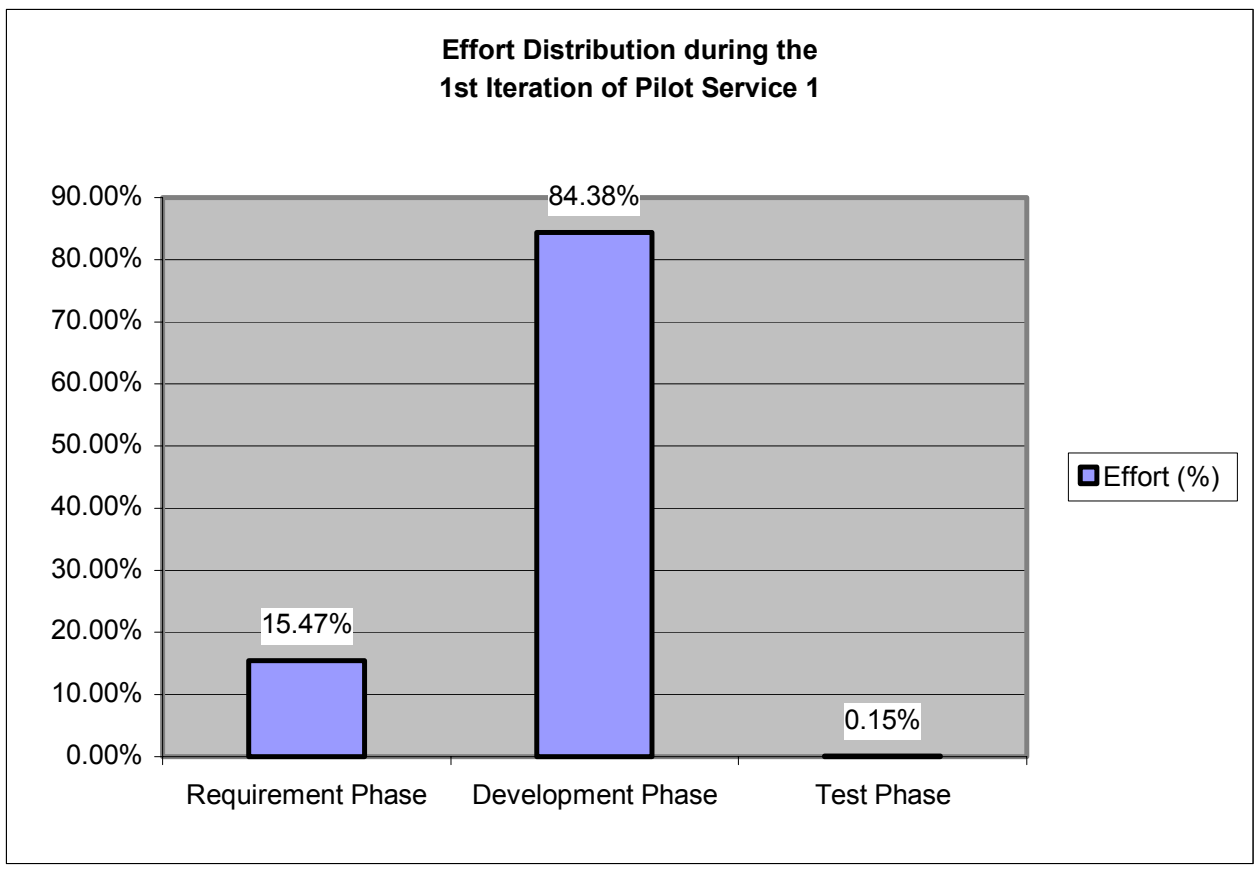

**Figure 4: Effort distribution, pilot service 1, iteration 1**

<u>Example</u> - The analysis of the quality models concerning effort and calendar time of pilot 1 after the first iteration revealed that, among other things, most of the effort as well as most of the time was spent on designing the technical infrastructure and on coding. Both activities belong to the development phase.

Figure 4 shows that about 85% of the effort was spent on the development phase. One interpretation given by the developers was that analysis and design were mainly concerned with technical issues (e.g., server support of WML, usability of the GUI for mobile devices, choice of a mobile suitable device).

In order to better address the analysis of the technical issues, a new activity called study feasibility and templates to document main feasibility issues were introduced into the requirements phase during the 2nd iteration. Figure 5 shows an excerpt of the modified EPG.

usability of the GUI for mobile devices, choice of a mobile suitable device).

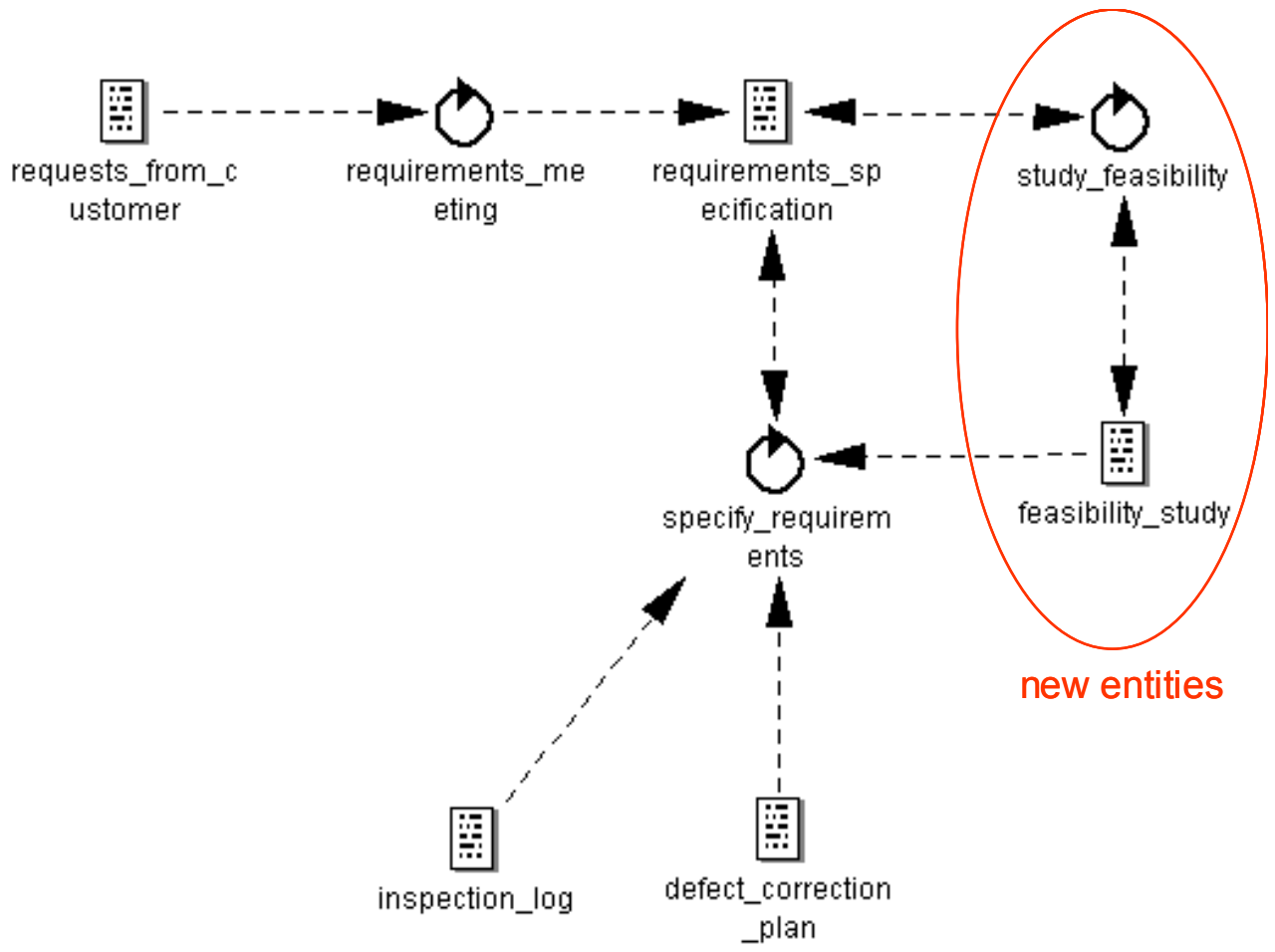

**Figure 5: Excerpt of the process model for the second iteration of pilot service 1**

As a result of the change, during iteration 2, the performers of pilot 1 produced a report documenting the main feasibility issues addressed. The performers were also able to focus the many technical issues more organically during the retrospectives and to provide a set of lessons learned regarding issues to be considered, for example, when porting an information system from a mobile phone to a PDA or to a BlackBerry device, or when transforming an information system based on WML into J2ME.

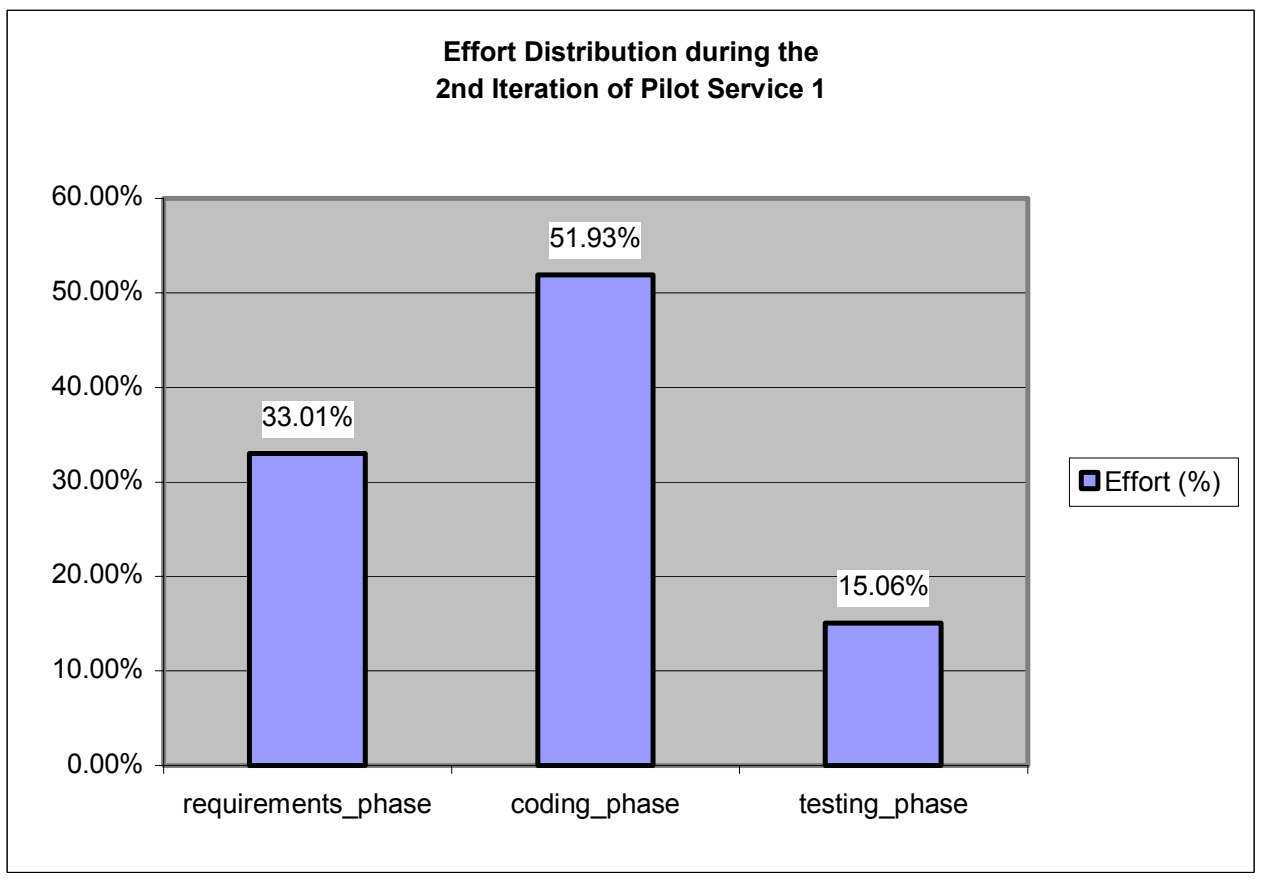

**Figure 6: Effort distribution, pilot service 1, iteration 2**

As a further consequence, during the second iteration, it became possible to distinguish how much of the development effort was spent on explorative work, as shown by Figure 6 where a proportionally greater amount of effort spent during the second iteration on the requirement phase indicates the effort spent on investigating feasibility issues.

*2) Improvement of Estimation Capabilities*

Description of the pilot - Pilot service 2 is concerned with the new development of a multi-player online game for mobile devices: many users interact in a shared environment, i.e., a virtual labyrinth. The players can collect different items, chat, and fight against enemies and against each other. From a business point of view, games and entertainment could be, after voice and SMS, the next killer application on the wireless Internet. The development is distributed between two different teams / organizations: one organization is responsible for the development of the client on the mobile device and provides a multimedia-messaging stack on the terminal part; the other organization customizes the multimedia layer on the server side.

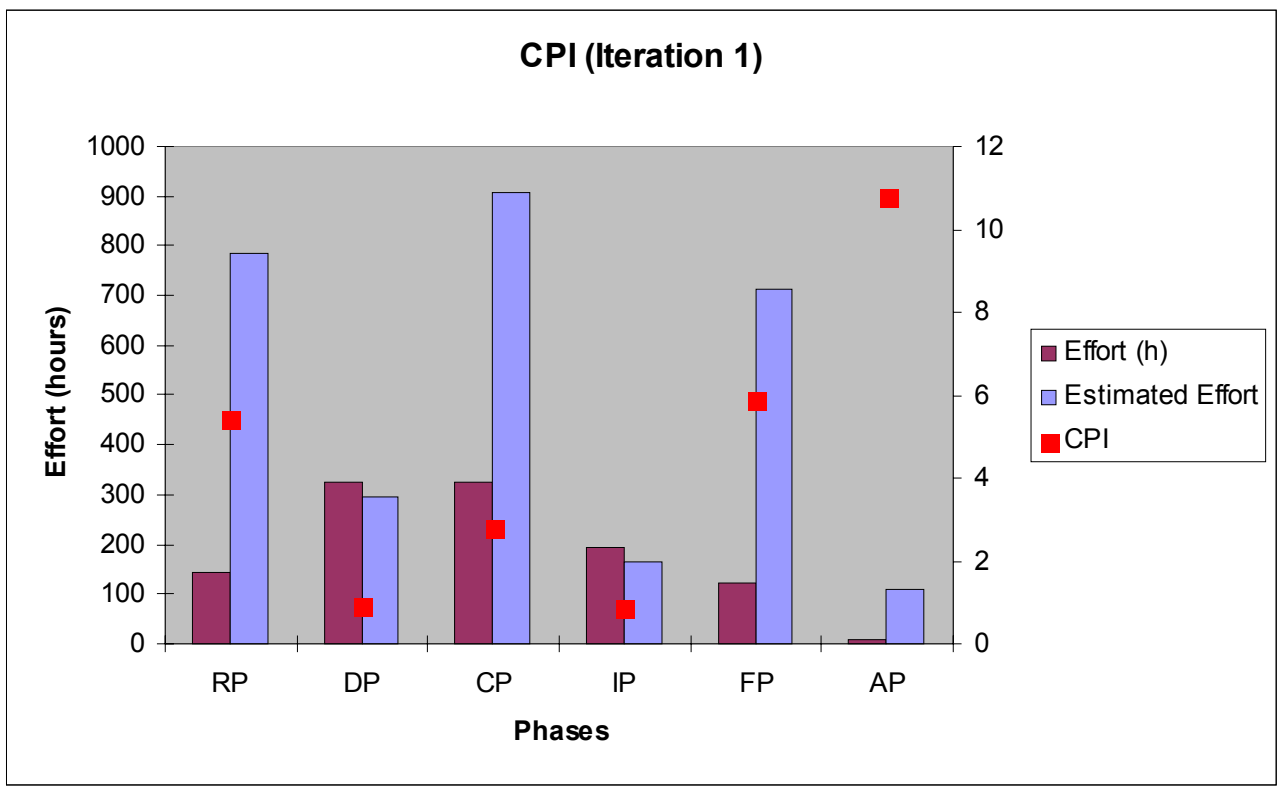

**Figure 7: Cost Performance Index, pilot service 2, iteration 1**

The organization responsible for the client side follows an iterative life cycle model consisting of four phases: requirements phase, design phase, coding phase, and testing phase. The process reaches CMM maturity level 3. The process is characterized by extensive use of verbal communication as well as explicit formal documentation.

Example – The Cost Performance Index (CPI = planned effort / actual effort, [6]) can be used as an indicator of the accuracy of the effort estimate. Figure 7 shows the CPI achieved for the different phases of development by the organization responsible for the client side of the pilot service 2 during the first iteration.

The average CPI of 4.45 and the greatest overestimation occurring in the acceptance phase, (AP, CPI of 10.8) indicate a systematic overestimation of effort.

In order to obtain more accurate estimates for the second iteration, the effort distribution data from the first iteration were used together with the first estimates as basis for the estimation process.

Figure 8 shows how the values for the new estimates were chosen from within a range between the data estimated before the beginning of the first iteration and the data gathered during the first iteration. Concerning the requirements phase, for example, approx. 26% was the estimate for the first iteration, 13% was the effort actually spent on this phase during the first iteration, and 17% was estimated for the second iteration. The new estimated value is less than the estimate from the first iteration, but greater than the value actually measured. The estimation values were also chosen according to the critical issues expected in the second iteration.

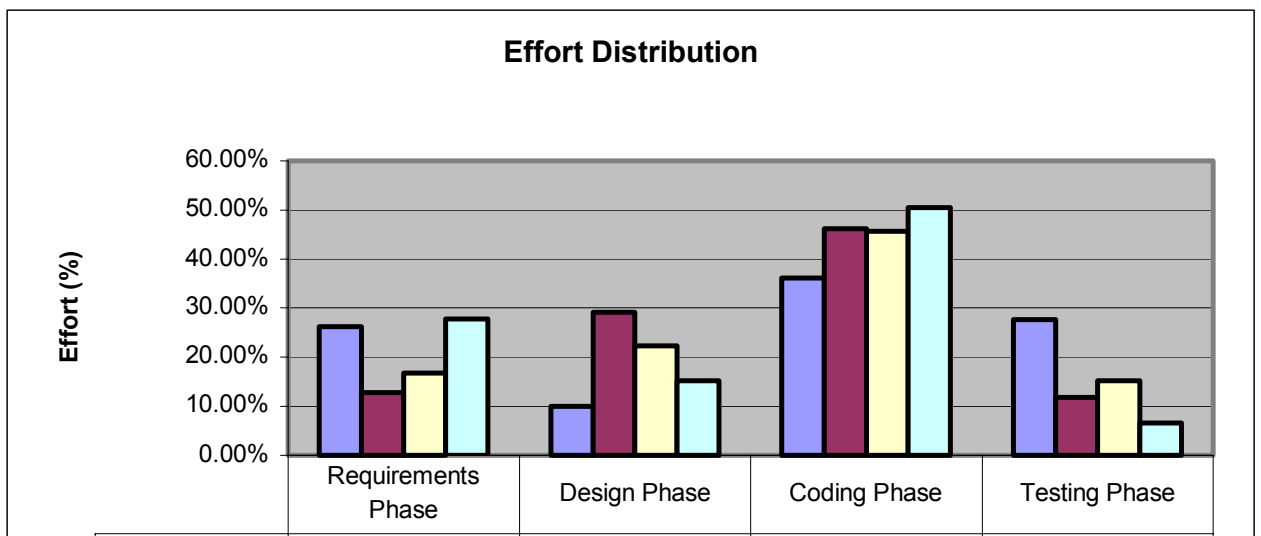

| | Requirements Phase | Design Phase | Coding Phase | Testing Phase |
|---|---|---|---|---|
| Estimate It. 1 | 26.31% | 9.98% | 36.09% | 27.62% |
| Effort It. 1 | 12.87% | 29.22% | 46.16% | 11.75% |
| Estimate It. 2 | 16.74% | 22.36% | 45.69% | 15.21% |
| Effort It. 2 | 27.74% | 15.19% | 50.52% | 6.55% |

Phases

**Figure 8: Overview of effort distributions and effort distribution estimates during the first two iterations**

A comparison of the CPI values achieved during the two iterations (see Figure 9), the average CPI of 2.06, the greatest overestimation (testing phase, CPI of 3.61), and the closest estimation (requirements phase, CPI of 0.94) indicate how the capability to provide more accurate effort estimates increased with the experience and the use of the effort baselines available from the first iteration.

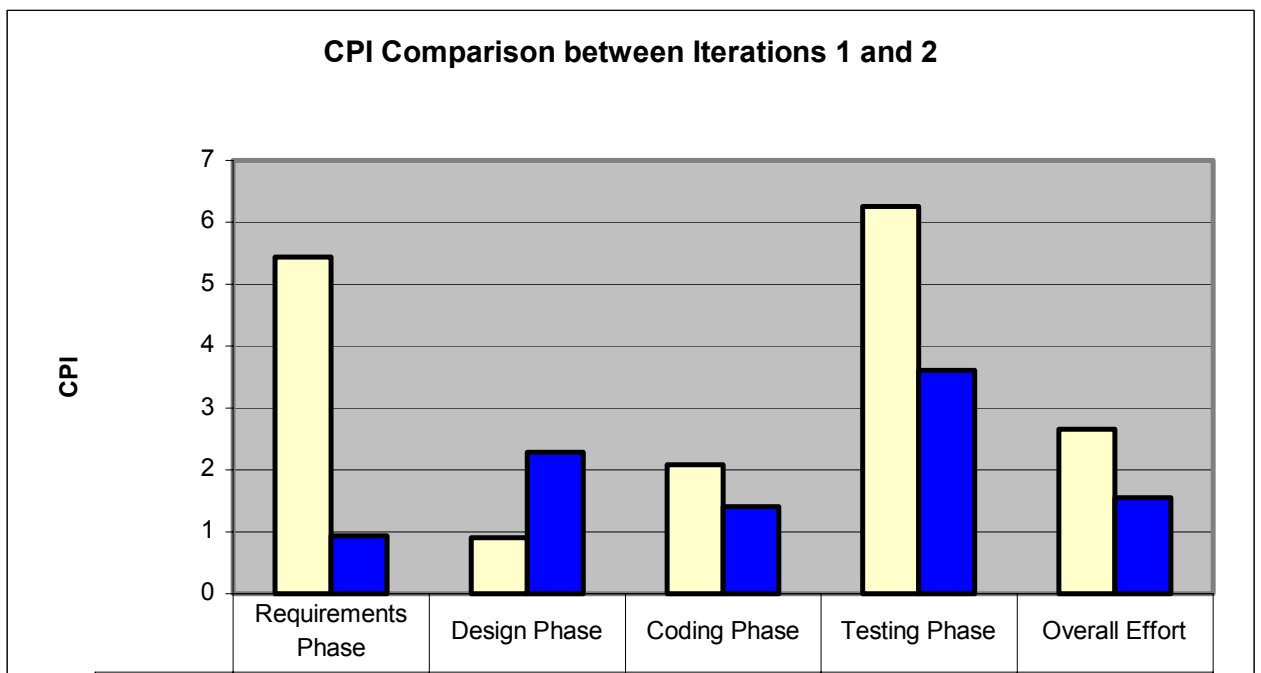

| | Requirements Phase | Design Phase | Coding Phase | Testing Phase | Overall Effort |
|---|---|---|---|---|---|
| CPI It. 1 | 5.4375 | 0.908256881 | 2.079380445 | 6.25095057 | 2.659517426 |
| CPI It. 2 | 0.937388193 | 2.287581699 | 1.404715128 | 3.606060606 | 1.553349876 |

**Figure 9: Comparison of the cost performance indices from the first two development iterations of pilot service 2 (client side)**

*3) Lessons Learned from Developing the Pilot Services*

The first issue to be considered when developing wireless Internet services is the great diversity of target devices in terms of display size and mode (i.e., resolution and number of colors), memory capacity, processor performance, and interaction mechanisms with the user (i.e., keyboard, joggle-wheel, cursor buttons, joystick, touch screen, voice control, etc.). This heterogeneity makes it very difficult to reconcile the need for portability with the increasing demand for appealing applications.

Even Java's promise of code working on every platform is difficult to achieve: different levels of compliance with the J2ME specification in the case of virtual machines implemented by different device manufacturers can lead to great variations in performance and behavior of the same application running on different mobile devices.

Another consideration concerns the maturity of the technologies specific to the wireless domain: many aspects of mobile devices (file system, network access capabilities, memory, etc.) are of much lower level than on regular desktop-systems. This has consequences in terms of predictability of the quality of services and the development process.

Technologies proven to be reliable when applied within the context of the traditional Internet may turn out to be unreliable or perform poorly when used within the context of the wireless world.

Testing wireless Internet services proved very challenging due to different reasons: The first reason are the many usability issues (e.g., consistent interfaces, navigation, access, etc.) related to the great diversity of devices available on the market. Most of the usability issues have to be further researched due to the novelty of the domain. Another reason is the development for future announced devices: Device specifications are subject to change without notice and are usually unreliable. Another reason is that a lot of effort has to be spent on setting a proper environment. Emulators represent one unsatisfactory but necessary alternative solution. The main advantage of using emulators is the automation of the testing procedures whereas unreliable behavior is their greatest disadvantage.

## IV. Conclusions

This article presented an approach for the development of software process baselines in new application domains that is based on the observation of representative pilots. Some results from the WISE project supported the description as concrete examples of the capabilities of the approach. In the following, some conclusive remarks are given.

The descriptive process modeling approach supported by the Spearmint® environment played a key role in stabilizing the processes and eliciting accurate process models. The accurate models provided the necessary basis for meaningful effort tracking and planning. Also, the online documentation of the development processes through EPGs proved to be very useful for disseminating process information and guiding process performers. Thus, the EPGs simplified cooperation between the different organizations involved in the highly distributed WISE project.

The development of software process baselines based on observations turned out to be easier to perform within organizations with a higher maturity level. Nonetheless, it could be performed successfully even within organizations characterized by the use of ad hoc processes.

As expected, effort estimation proved to be a challenging process. During the first iteration, the organizations involved were not able to deliver effort estimates or the estimates they delivered turned out to be inaccurate at the end of the iteration. On the other hand, effort tracking performed during the first iteration together with estimation processes based on the effort data collected led to much more accurate effort estimates in the second iteration.

Also, the great role played by qualitative data in providing an insight into the emerging wireless Internet application domain should be pointed out: Lessons learned provided an insight into details of the development processes that had been difficult to capture when only quantitative data were collected. At the same time, lessons learned proved to be very helpful in interpreting quantitative data.

## Acknowledgment

We would like to thank the WISE consortium, especially the pilot partners, for their fruitful cooperation. We would also like to thank Sonnhild Namingha, from the Fraunhofer Institute for Experimental Software Engineering (IESE) for reviewing the first version of the article.